\documentclass{llncs}

\let\llncssubparagraph\subparagraph
\let\subparagraph\paragraph
\usepackage[compact]{titlesec}
\let\subparagraph\llncssubparagraph

\usepackage{graphicx}  
\usepackage{amsfonts}
\usepackage{booktabs}
\usepackage{siunitx}
\usepackage{multirow}
\usepackage{xcolor}
\usepackage{pifont} 
\usepackage{threeparttable}
\usepackage{tabularx}
\usepackage{caption}
\usepackage{hyperref}
\hypersetup{colorlinks=false,citecolor=green, final}

\definecolor{grey}{rgb}{0.33, 0.33, 0.33}
\definecolor{red}{rgb}{1,0,0}
\definecolor{blue}{rgb}{0,0,1}

\newcolumntype{C}{>{\centering\arraybackslash}X}

\newcommand\mdoubleplus{\mathbin{+\mkern-10mu+}} 

\newcommand{\spademark}{\ding{68}}%
\newcommand{\tmark}{\ding{61}}%

\begin{document}
\frontmatter          
\pagestyle{headings}  
%
%
\mainmatter              
\title{Image Super Resolution via Bilinear Pooling: Application to Confocal Endomicroscopy}
\titlerunning{Super Resolution for Confocal Endomicroscopy}  
%

%
\authorrunning{Izadi et al.} 
%

%

\author{Saeed Izadi,  Darren Sutton, and Ghassan Hamarneh}
\institute{
School of Computing Science, Simon Fraser University, Canada \\
\email{\{saeedi, darrens, hamarneh\}@sfu.ca}\\
}

\maketitle              

\begin{abstract}
Recent developments in image acquisition literature have miniaturized the confocal laser endomicroscopes to improve usability and flexibility of the apparatus in actual clinical settings. However, miniaturized devices collect less light and have fewer optical components, resulting in pixelation artifacts and low resolution images. Owing to the strength of deep networks, many supervised methods known as super resolution have achieved considerable success in restoring low resolution images by generating the missing high frequency details. In this work, we propose a novel attention mechanism that, for the first time, combines 1st- and 2nd-order statistics for pooling operation, in the spatial and channel-wise dimensions. We compare the efficacy of our method to 10 other existing single image super resolution techniques that compensate for the reduction in image quality caused by the necessity of endomicroscope miniaturization. All evaluations are carried out on three publicly available datasets. Experimental results show that our method can produce superior results against state-of-the-art in terms of PSNR, and SSIM metrics. Additionally, our proposed method is lightweight and suitable for real-time inference.  
\end{abstract}

\section{Introduction}
Colorectal cancer is known as the fourth most-common cancer and remains one of the leading causes of cancer related mortality in the world. In 2018, more than 1 million people were affected by colorectal cancer worldwide, resulting in an estimated 550,000 deaths~\cite{bray2018global}. 
Rapid histopathologic assessment is an important tool that may improve disease prognosis by detecting early-stage cancer and pre-cancerous conditions. Although biopsy and \textit{ex-vivo} tissue examination are widely accepted as the diagnostic gold standard, such procedures take time and may limit the ability of the endoscopist to rapidly gauge disease severity. Confocal laser endomicroscopy (CLE), on the other hand, has substantially improved real-time \textit{in-vivo} visualization of the subsurface of living cells, vascular structures, and tissue patterns during endoscopic examination~\cite{kiesslich2004confocal}. \par

For \textit{in-vivo} histological examination, the large size of the microscope complicates navigation of the interior of the body in a clinical setting. Therefore, it is necessary to reduce the size of the microscope to completely and safely access the organ(s) of interest. However, miniaturization reduces the number of optical elements in the microscope probe, introducing pixelation artifacts in the acquired images. One strategy to remove image artifacts and enhance image quality is to directly post-process degraded images. An emerging process in the field of image processing, referred to as single image super-resolution (SR), aims to reconstruct an accurate high-resolution (HR) image given its low-resolution (LR) counterpart. Thus, SR is a promising software method to mitigate image degradation due to hardware miniaturization.

Among traditional SR algorithms, Huang et al.~\cite{Huang2015} proposed leveraging self-similarity modulo affine transformations to accommodate natural deformation of recurring statistical priors within and across scales of an image. Timofte et al.~\cite{Timofte2013,Timofte2014} used a combination of neighbour embedding and sparse dictionary learning over an external database and proposed anchored neighborhood regression in the dictionary atom space.  Recently, CNNs have advanced the SR research field by directly learning the mapping between LR and HR images~\cite{Dong2014,Kim2016,KimDRCN2016,Lai_LapSRN_2018,Ahn_2018_ECCV}. Dong et al.~\cite{Dong2014} demonstrated that a fully convolutional network trained end-to-end can perform LR-to-HR nonlinear mapping. Kim et al.~\cite{Kim2016} suggested a trained network to predict additive details in the form of a residual image, which is summed with the interpolated image. Kim et al.~\cite{KimDRCN2016} addressed model overfitting by reducing the number of parameters via recursive convolutional layers. Lai et al.~\cite{Lai_LapSRN_2018} designed a network which progressively reconstructs the sub-band residuals of high-resolution images at multiple pyramid levels. Ahn et al.~\cite{Ahn_2018_ECCV} improved speed and efficiency of SR models by designing a cascade mechanism over residual networks. Lastly, Cheng et al.~\cite{Cheng_sesr_2018} exploited recursive squeeze and excitation modules in a network to exploit relationships between channels. Izadi et al.~\cite{Izadi2018} reported the first attempt to deploy CNNs on CLE images. They used a densely connected CNN to transform synthetic LR images into HR ones. Ravi et al.~\cite{ravi2018effective} employed a CNN to restore missing details into LR images. They collected a set of consecutive LR frames and generated synthetic HR images using a video registration technique. 
In a more recent study~\cite{RAVI2019123}, Ravi et al. trained a CNN for unsupervised SR on CLE images using a cycle consistency regularization, designed to impose acquisition properties on the super-resolved images. \par 

In this paper, we present a lightweight convolutional neural network (CNN) that is appropriate for frame-wise SR by incorporating a novel attention mechanism. In contrast to SESR~\cite{Cheng_sesr_2018}, which leverages attention modules from the Squeeze-and-Excitation network (SENet)~\cite{Hu_2018_CVPR} to re-weight channels, we introduce a novel weighting scheme to recalibrate learned features based on pairwise relationships. Our attention modules compromise both 1$^{\text{st}}$-order pooling and 2$^{\text{nd}}$-order pooling (a.k.a. bilinear pooling), improving the quality of learned features in the network by considering pairwise correlations along feature channels and spatial regions~\cite{gao2018global}. The compactness and computational speed of our network lends well to real-time implementation during \textit{in-vivo} examination. We demonstrate that stacking attention modules in the middle of a low-level feature extraction head and a feature integration tail quantitatively and qualitatively produces superior results against existing SR methods and generalizes well over unseen microscopic datasets.


\begin{figure*}[t!]
\centering
  \includegraphics[width=\textwidth]{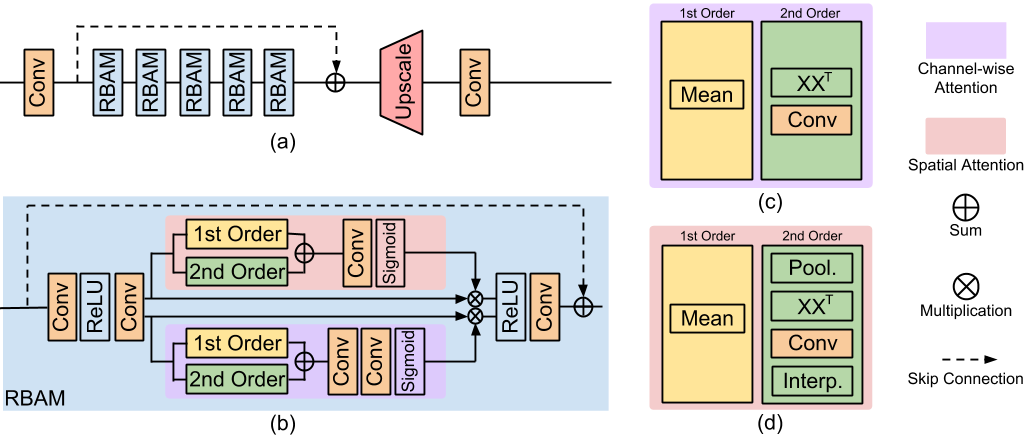}
  \caption{(a) The overall architecture of our proposed network. (b) RBAM architecture. (c) channel-wise and (d) spatial attention architectures.}
  \label{fig:overall_arch}
\vspace{-3mm}
\end{figure*}

\section{Method}

\noindent \textbf{Network Overview.}
Fig.~\ref{fig:overall_arch}-a depicts the overall architecture of our proposed LR-to-SR network. Let $\mathrm{L}_{\mathrm{LR}} \in \mathbb{R}^{1 \times W \times H}$, $\mathrm{I}_{\mathrm{SR}} \in \mathbb{R}^{1 \times rH \times rW}$, and $r$ denote the low resolution input and super-resolved output, and the downsample factor, respectively. We use a convolution layer, denoted by $\mathcal{F}(\cdot)$, with a $3 \times 3$ kernel and $C$ output channels to extract initial features $\mathrm{H}^{\mathrm{0}} \in \mathbb{R}^{C \times H \times W}$ , i.e.
\begin{equation}
   \mathrm{H}^{\mathrm{0}} = \mathcal{F}_{\mathrm{c}}(\mathrm{I}_{\mathrm{LR}};\theta_{0}),
\end{equation}
where $\theta$ refers to the learnable parameters. In our proposed network, the initial features $\mathrm{H}^{\mathrm{0}}$ are updated by sequential residual attention modules, denoted as $\mathcal{G(\cdot)}$ and a skip connection. The entire high-level feature extraction stage is denoted as $\mathcal{B}(\cdot)$: 
\begin{equation}
   \mathrm{H}^{\mathcal{B}} = \mathcal{B}(\mathrm{H}^{\mathrm{0}}) = \mathcal{G}^{\mathrm{b}}(\mathcal{G}^{\mathrm{b-1}}(...(\mathcal{G}^{\mathrm{1}}(\mathrm{H}^{\mathrm{0}}))...)) + \mathrm{H}^{\mathrm{0}}.
\end{equation}
To upsample the feature maps, we use sub-pixel convolutions, denoted as $\mathcal{U}(\cdot)$, followed by a single channel $1 \times 1$ convolution for SR reconstruction:
\begin{equation}
  \mathrm{I}_{\mathrm{SR}} = \mathcal{F}_{\mathrm{1}}(\mathcal{U}(\mathrm{H}^{\mathcal{B}};\theta_{up}); \theta_{rec}).
  \vspace{-3mm}
\end{equation}

\noindent \textbf{Residual Bilinear Attention Module.}
In our proposed RBAM, we combine 1$^{\text{st}}$- and 2$^{\text{nd}}$-order pooling operations spatially and channel-wise to recalibrate learned features for efficient network training.
Fig.~\ref{fig:overall_arch}-b illustrates the structure of our proposed RBAM. Mathematically, we formulate RBAM as: 
\begin{equation}
  \mathrm{H}^{\mathrm{b}} = \mathcal{G}^{\mathrm{b}}(\mathrm{H}^{\mathrm{b-1}}) = \mathcal{Q}^{\mathrm{b}}(\mathrm{H}^{\mathrm{b-1}}) + \mathrm{H}^{\mathrm{b-1}},
\end{equation}
where $\mathcal{Q}(\cdot)$ denotes the attention modules before the skip connection. Given the input feature maps $\mathrm{H^\mathrm{b}} \in \mathbb{R}^{C \times H \times W}$, two convolutions with $3 \times 3$ kernel size interleaved with a ReLU activation function are performed to produce high-level feature maps $\mathrm{H^\mathrm{b}_{\mathrm{conv}}} \in \mathbb{R}^{C \times H \times W}$ as input to the attention branches:
\begin{equation}
  \mathrm{H}^{\mathrm{b}}_{\mathrm{conv}} = \mathcal{F}_{\mathrm{c}}(\mathcal{F}_{\mathrm{c}}(\mathrm{H}^{\mathrm{b-1}};\theta_{1}^{b});\theta_{2}^{b}).
\end{equation}

\noindent \textbf{Channel-wise Attention (CA) Branch.} CA leverages the inter-channel correspondence between feature responses (Fig.~\ref{fig:overall_arch}-c). 1$^{\text{st}}$- and 2$^{\text{nd}}$-order pooling mechanisms operate on $\mathrm{H}^{\mathrm{b}}_{\mathrm{conv}}$, producing two vectors $\mathrm{F_\mathrm{ca}^\mathrm{1st}}$, $\mathrm{F_\mathrm{ca}^\mathrm{2nd}} \in \mathbb{R}^{C \times 1 \times 1}$.
$\mathrm{F_\mathrm{ca}^\mathrm{1st}}$ is the 
1$^{\text{st}}$-order CA obtained by spatial average pooling to squeeze the feature map of each channel~\cite{Hu_2018_CVPR}. 
To obtain 2$^{\text{nd}}$-order CA, pairwise channel correlations are computed in the form of a covariance matrix $\mathrm{\mathrm{\Sigma}} \in \mathbb{R}^{C \times C}$ by spatial flattening, dimension permutation, and matrix multiplication. Each row in $\mathrm{\Sigma}$ encodes the statistical dependency of a channel with respect to every other channel~\cite{gao2018global}. Given the covariance matrix $\mathrm{\Sigma}$, we adopt a row-wise convolution with $1 \times C$ kernel size to produce the 2$^{\text{nd}}$-order CA vector $\mathrm{F_\mathrm{ca}^\mathrm{2nd}}$. Finally, two successive 1-D convolutions interleaved with a ReLU activation function operate on a vector formed by the sum of $\mathrm{F_\mathrm{ca}^\mathrm{1st}}+\mathrm{F_\mathrm{ca}^\mathrm{2nd}}$. The output of the convolution operation is fed into a sigmoid function $\sigma$, followed by element-wise multiplication $\otimes$ to produce the \textit{b}$^{th}$ updated features maps 
$\mathrm{H}_{\mathrm{ca}}^{\mathrm{b}}$:
\vspace{-0.5mm}
\begin{equation}
  \mathrm{H}_{\mathrm{ca}}^{\mathrm{b}} = \mathrm{H}^{\mathrm{b}}_{\mathrm{conv}} \otimes \sigma(\mathcal{F}_{\mathrm{c}}(\mathcal{F}_{\frac{\mathrm{c}}{4}}(\mathrm{F_\mathrm{ca}^\mathrm{1st}} + \mathrm{F_\mathrm{ca}^\mathrm{2nd}};\theta_{3}^{b});\theta_{4}^{b})).
\end{equation}


\noindent \textbf{Spatial Attention (SA) Branch.} SA indicates shared correspondence between spatial regions across all feature maps (Fig.~\ref{fig:overall_arch}-d). Given $\mathrm{H}^{\mathrm{b}}_{\mathrm{conv}}$ as the input, the 1$^{\text{st}}$-order spatial attention matrix, $\mathrm{F}_{\mathrm{sa}}^{\mathrm{1st}} \in \mathbb{R}^{1 \times H \times W}$, is computed by the average pooling operation along channel dimension to aggregate information for each spatial location across all features. To compute 2$^{\text{nd}}$-order spatial attention matrix, $\mathrm{F}_{\mathrm{sa}}^{\mathrm{2nd}}\in \mathbb{R}^{1 \times H \times W}$, we first reduce the spatial size of feature maps to $H' \times W'$ ($8 \times 8$ in our implementation) by applying average pooling. Then, appropriate reshaping, dimension permutation and matrix multiplication is adopted to obtain the covariance matrix $\mathrm{\Sigma} \in \mathbb{R}^{H'W' \times H'W'}$. Similar to channel-wise attention, a row-wise convolution with $1 \times H'W'$ kernel size is applied on $\mathrm{\Sigma}$. Eventually, dimension permutation and nearest neighbor interpolation produce $\mathrm{F}_{\mathrm{sa}}^{\mathrm{2nd}}$. We add these two matrices together element-wise and apply a convolution with $1 \times 1$ kernel size that feeds a sigmoid function. Spatial attention is realised by element-wise multiplication over all feature maps, formulated as:
\begin{equation}
  \mathrm{H}_{\mathrm{sa}}^{\mathrm{b}} = \mathrm{H}^{\mathrm{b}}_{\mathrm{conv}} \otimes \sigma(\mathcal{F}_{\mathrm{c}}(\mathrm{F_\mathrm{sa}^\mathrm{1st}} + \mathrm{F_\mathrm{sa}^\mathrm{2nd}};\theta_{5}^{b}))
\end{equation}

\begin{figure*}[t!]
\centering
\includegraphics[width=\textwidth]{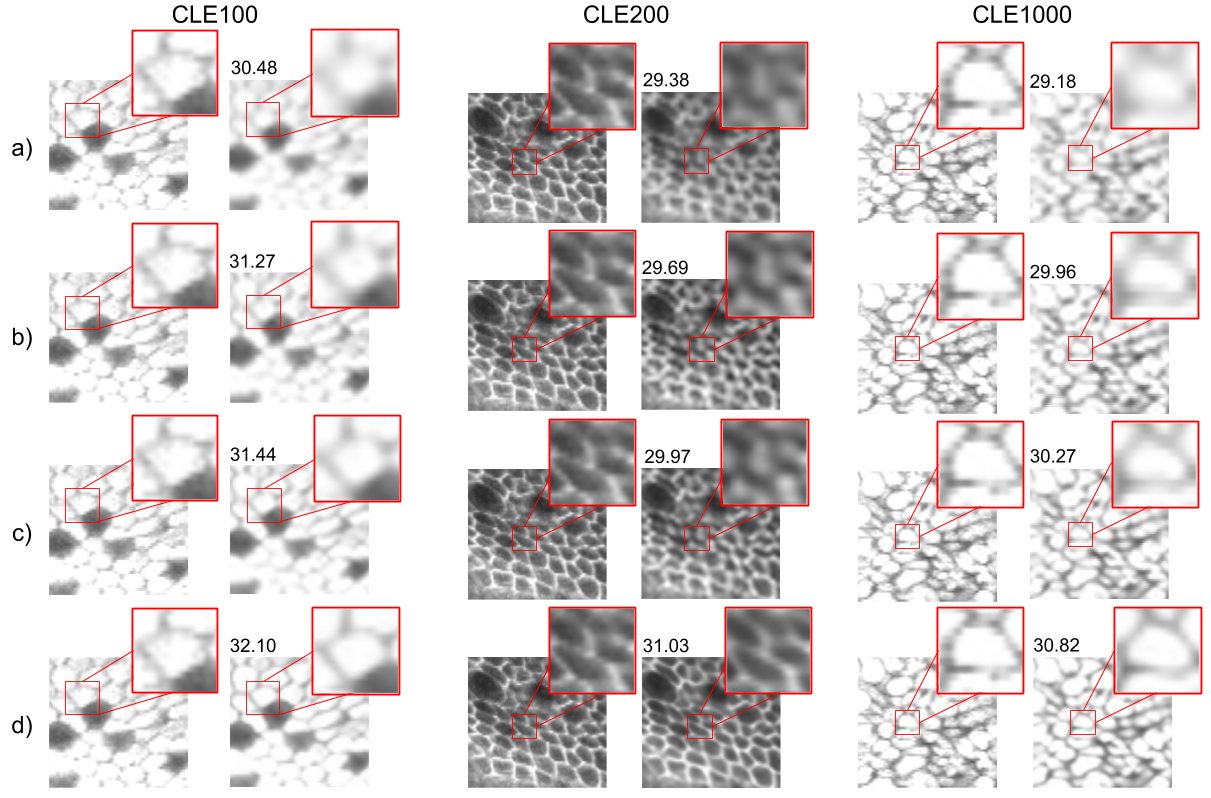}
\caption{Qualitative results and their PSNR scores at $\times 4$ SR. Each row shows the side-by-side comparison of HR with a) bicubic, b) GR, c) SESR, and d) RBAM across three datasets. HR images are shown for each pair for ease of visual comparison. }
\label{fig:qualitative}
  
\vspace{-3mm}
\end{figure*}

\noindent \textbf{Attention Fusion.} The updated features are concatenated ($\mdoubleplus$) and aggregated via a convolution with kernel $1 \times 1$ kernel. Lastly, $\mathrm{H}^{\mathrm{b}}$ is added via skip connection:
\begin{equation}
  \mathrm{H}^{\mathrm{b}} = \mathcal{F}_{\mathrm{c}}(\mathrm{H}^{\mathrm{b}}_{\mathrm{ca}}  \mdoubleplus  \mathrm{H}^{\mathrm{b}}_{\mathrm{sa}};\theta_{6}^{b}) + \mathrm{H}^{\mathrm{b-1}}.
\end{equation}

\section{Results and Discussion}
\begin{figure*}[t!]
\centering
  \includegraphics[width=\textwidth]{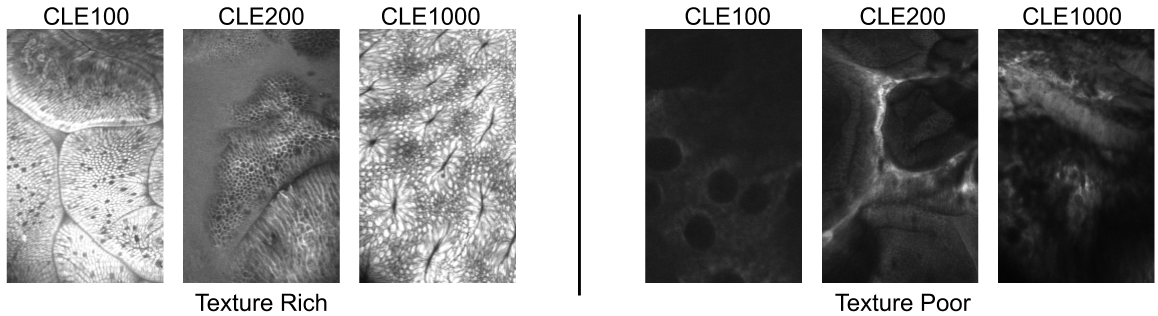}
  \caption{(a) Examples of images from the partitioned test set. Images belonging to the 'texture rich' partition are used for evaluation.}
  \label{fig:dataset}
\vspace{-3mm}
\end{figure*}

\noindent \textbf{Data}. We evaluate existing state-of-the-art SR methods, as well as our proposed RBAM, on three publicly available CLE datasets (Table \ref{tab:datasets}). We select images rich in texture by assessing the SR performance of bicubic interpolation on the unseen test set. As depicted in Fig.~\ref{fig:dataset}, images with PSNR scores below the mean PSNR score of the bicubic method evalulated on the test set are deemed 'texture rich', and are used for evaluation, whereas images associated with scores above the mean are deemed 'texture poor'. In other words, images which can be effectively restored using bicubic interpolation are rejected for evaluation, as they contain little information on which to assess the performance of state-of-the-art methods. Evaluation assesses the methods' ability to reconstruct $1024 \times 1024$ HR image from a synthesized LR counterpart obtained via bicubic downsampling with the appropriate factor ($\times$2 or $\times$4). \par

\begin{table}[!t]
    \centering
    \caption{Details of the datasets used in our evaluation.}
    \label{tab:data}
    \begin{tabular}{|l|l|c|c|c|c|}
        \hline
        dataset&provided by&\#patients& \#images & anatomical site&image size \\
         \hline
         \hline
         CLE100&Leong et al.~\cite{cle100}&30&181&small intestine&$1024 \times 1024$\\
         \hline
         CLE200&Grisan et al.~\cite{cle200}&32&262&esophagus&$1024 \times 1024$\\
         \hline
         CLE1000&{\c{S}}tef{\u{a}}nescu et al.~\cite{cle1000}&11&1025&colorectal mucosa&$1024 \times 1024$\\
         \hline
    \end{tabular}
    \label{tab:datasets}
    \vspace{-3mm}
\end{table}

\noindent \textbf{Training Settings}. We train all methods on a random partition (80\%) of CLE100, and evaluate them on the remaining 20\%  as well as CLE200, and CLE1000. 
For DL-based methods, we replicated the reported training settings, and used public code for traditional algorithms. For our model, we use $B=5$ RBAMs and set the number of features to $C=64$ to create a lightweight network. In each training batch, 16 LR patches of size $48 \times 48$ are randomly extracted as inputs, and augmented by random 90$^{\circ}$ rotations and horizontal/vertical flip. 
We use Adam optimizer and L1 loss to train our network for 300 epochs.
Initial learning rate is set to $10^{-4}$ and is halved every 50 epochs. \par

\noindent \textbf{Ablation Investigation}. We discern the effectiveness of the individual components in our network modules by ablating attention blocks and evaluating performance after 50 epochs. Our investigation shows that, for CLE100 at $\times 2$ SR, attention-based variants outperform the baseline, demonstrating the merits of incorporating spatial and channel-wise contextual information. We also observed that using both 1$^{\text{st}}$ and 2$^{\text{nd}}$-order pooling operations simultaneously outperform using either 1$^{\text{st}}$ or 2$^{\text{nd}}$-order channel-wise pooling individually. We similarly note that using both spatial and channel-wise attention outperforms either one alone. \par

\noindent \textbf{Comparison to State-of-the-art.} We compare the performance of traditional algorithms including ANR~\cite{Timofte2013}, GR~\cite{Timofte2013} and A+~\cite{Timofte2014}, as well as DL-based techniques including SRCNN~\cite{Dong2014}, VDSR~\cite{Kim2016}, DRCN~\cite{KimDRCN2016}, LapSRN~\cite{Lai_LapSRN_2018}, SESR~\cite{Cheng_sesr_2018} and our proposed RBAM. Table~\ref{tab:scores} summarizes the quantitative comparisons in terms of peak signal-to-noise-ratio (PSNR-SEM), structural similarity (SSIM), and inference time at $\times 2$, and $\times 4$ SR. From the table, one can see that most DL-based methods consistently outperform traditional SR algorithms in PSNR and SSIM metrics. Particularly, RBAM significantly outperforms the mean PSNR over all datasets by 0.18dB and 0.13dB for $\times 2$ and $\times 4$ SR, respectively. Furthermore, RBAM is a practical compromise between inference time, and generalization. Our results show a moderate quantitative increase in PSNR score and a considerable increase in qualitative performance - this is similar to previous works in single image super resolution~\cite{yang2019deep}. Fig.~\ref{fig:qualitative} shows selected image patches from each dataset for qualitative assessment. RBAM can delicately restore high-frequency cues, such as granular textures and sudden changes in grayscale pixel intensity. This manifests qualitatively in the form of improved restoration of high frequency details such as cell membranes (CLE200, CLE1000 examples) and intracellular spaces (CLE100 example). \par
\noindent \textbf{Motivation for Bilinear Pooling.}
We combine 1$^{\text{st}}$-order and 2$^{\text{nd}}$-order pooling to recalibrate learned features based on channels that activate often or correspond to feature rich inputs, respectively. Channels that activate often are likely responding to common, low frequency image features. Conversely, channels that are highly correlated may be responding to feature rich instances in the image space that activate multiple filters simultaneously. High frequency features tend to be complex, and not as common semantically compared to low frequency image details. Therefore, channels that learn complex image features may not be emphasized by first order pooling operations alone. Combining first and second order pooling in an attention module assures that hard working channels are rewarded without diminishing the optimization of channels that learn complex features in the low to high resolution image mapping space.   \par
\begin{table}[t!]
\scriptsize
\renewcommand{\arraystretch}{0.7}
\begin{threeparttable}
\centering
\caption{Quantitative results of SR models at $\times 2$ and $\times 4$ factors. \textbf{Bold} indicates the best result. \text{\spademark} and \text{\tmark} denote traditional and DL-based methods, respectively. PSNR scores are reported with the standard error of the mean (SEM) for each method.}
\label{tab:scores}
\begin{tabular*}{\textwidth}{l
    @{\extracolsep{\fill}}cccccccrcc}
    \toprule 
    Methods & & 
    \multicolumn{2}{c}{CLE100} & \multicolumn{2}{c}{CLE200} & \multicolumn{2}{c}{CLE1000} & 
    time \\
    
    \cmidrule(l){3-4} \cmidrule(l){5-6} \cmidrule(l){7-8}

    \footnotesize Scale $\times$2 & & {\scriptsize PSNR} & {\scriptsize SSIM}  & {\scriptsize PSNR} & {\scriptsize SSIM} &  {\scriptsize PSNR} & {\scriptsize SSIM} &\\
    
    \cmidrule(r){1-1} \cmidrule(l){3-4} \cmidrule(l){5-6} \cmidrule(l){7-8}
    \cmidrule(r){1-1}

    \scriptsize Bicubic  &
    & \text{33.69$\pm$0.06} & \text{0.8693} & \text{35.53$\pm$0.01} & \text{0.9029} &  \text{34.45$\pm$0.01}   & \text{0.8920}  & \text{0.02}\\
    
    
    \scriptsize A+\tnote{\spademark}~\cite{Timofte2014}&
    & \text{34.22$\pm$0.07} & \text{0.8928}  & \text{36.14$\pm$0.01} & \text{0.9218}  &  \text{35.04$\pm$0.01} & \text{0.9114} &  \text{6.72}\\
    
    \scriptsize ANR\tnote{\spademark}~\cite{Timofte2013} &
    & \text{36.44$\pm$0.13}  & \text{0.9226}      & \text{39.10$\pm$0.01}  & \text{0.9559} &  \text{37.64$\pm$0.01}    &\text{0.9559}    &\text{6.07}\\
    
    \scriptsize GR\tnote{\spademark}~\cite{Timofte2013} &
    & \text{36.56$\pm$0.13}  & \text{0.9243}    & \text{39.26$\pm$0.01} & \text{0.9579} & \text{37.79$\pm$0.01}    &\text{0.9448}        &\text{4.47}\\
    
    \scriptsize SRCNN\tnote{\tmark}~\cite{Dong2014} &
    & \text{35.75$\pm$0.11} & \text{0.9181}  & \text{38.25$\pm$0.01} & \text{0.9494}  &   \text{36.87$\pm$0.01}   & \text{0.9380}  & \text{0.06}\\
    
    \scriptsize VDSR\tnote{\tmark}~\cite{Kim2016} &
    & \text{36.72$\pm$0.13}             & \text{0.9276}                  & \text{39.31$\pm$0.01}           & \text{0.9578}        &   \text{37.89$\pm$0.01} & \text{0.9462}  & \text{0.25}\\
    
    \scriptsize DRCN\tnote{\tmark}~\cite{KimDRCN2016} &
    & \text{36.65$\pm$0.13}             & \text{0.9257}                   & \text{39.29$\pm$0.01}           & \text{0.9575}         &   \text{37.83$\pm$0.01} & \text{0.9452}  & \text{0.48}\\
    
    \scriptsize LapSRN\tnote{\tmark}~\cite{Lai_LapSRN_2018} &
    & \text{36.71$\pm$0.13}             & \text{0.9264}                  & \text{39.25$\pm$0.01}           & \text{0.9583}                   &   \text{37.91$\pm$0.01} & \text{0.9462}  & \text{0.07}\\
    
    \scriptsize SESR\tnote{\tmark}~\cite{Cheng_sesr_2018} &
    & \text{36.76$\pm$0.13}             & \text{0.9282}                 & \text{39.36$\pm$0.01}           & \text{0.9583}       
    & \text{37.91$\pm$0.01} & \text{0.9462}  & \text{0.27}\\
    
    \scriptsize RBAM (Ours)\tnote{\tmark} &
    & \textbf{36.91$\pm$0.12} & \textbf{0.9321}  & \textbf{39.45$\pm$0.01}             & \textbf{0.9590}  &   
    \textbf{38.22$\pm$0.01} & \textbf{0.9501} & \text{0.18}\\
    
    \addlinespace
    \footnotesize Scale $\times$4 &&&&  &&&&\\
    \cmidrule(r){1-1}
    \cmidrule(r){1-1}
    
    \scriptsize Bicubic & 
    & \text{31.29$\pm$0.04} & \text{0.6673} &   \text{32.45$\pm$0.01} & \text{0.7318}  &   \text{31.78$\pm$0.01} &\text{0.7278}  & \text{0.02}\\
    
    
    \scriptsize A+~\cite{Timofte2014}&
    & \text{31.57$\pm$0.04} & \text{0.7042}    & \text{32.76$\pm$0.01} & \text{0.7607} &    \text{32.06$\pm$0.01} & \text{0.7517}  & \text{3.03}\\
    
    \scriptsize ANR~\cite{Timofte2013} &
    & \text{31.68$\pm$0.04} & \text{0.7160}  & \text{32.93$\pm$0.01} & \text{0.7736} &  \text{32.23$\pm$0.01}  & \text{0.7671}  & \text{2.88}\\
    
    \scriptsize GR~\cite{Timofte2013}&
    & \text{31.70$\pm$0.04} & \text{0.7201}  & \text{32.95$\pm$0.01} & \text{0.7736} &  \text{32.25$\pm$0.01}  & \text{0.7703} & \text{2.31}\\
    
    \scriptsize SRCNN~\cite{Dong2014}&
    & \text{31.59$\pm$0.04}          & \text{0.7073}    &   \text{32.76$\pm$0.01} & \text{0.7617}  &   \text{32.07$\pm$0.01} & \text{0.7566}  & \text{0.06} \\
    
    \scriptsize VDSR~\cite{Kim2016}&
    & \text{31.66$\pm$0.04}          & \text{0.7144}   &   \text{32.86$\pm$0.01} & \text{0.7804} &   \text{32.16$\pm$0.01} & \text{0.7635}  & \text{0.25}\\
    
    \scriptsize DRCN~\cite{KimDRCN2016}&
    & \text{31.70$\pm$0.04}          & \text{0.7214}   &   \text{32.92$\pm$0.01} & \text{0.7750} &    \text{32.21$\pm$0.01} & \text{0.7635}  & \text{0.48}\\
    
    \scriptsize LapSRN~\cite{Lai_LapSRN_2018}&
    & \text{31.68$\pm$0.04}          & \text{0.7190}    & \text{32.76$\pm$0.01}  & \text{0.7617} &  \text{32.29$\pm$0.01} & \text{0.7737}  &  \text{0.08}\\
    
    \scriptsize SESR~\cite{Cheng_sesr_2018}&
    & \text{31.76$\pm$0.04}          & \text{0.7249}   &   \text{32.99$\pm$0.01} & \text{0.7804} &   \text{32.29$\pm$0.01} & \text{0.7737} & \text{0.33}\\
    
    \scriptsize RBAM (Ours) &
    & \textbf{31.84$\pm$0.04} & \textbf{0.7315}  &   \textbf{33.11$\pm$0.01} & \textbf{0.7852}  &   \textbf{32.47$\pm$0.01} & \textbf{0.7874} & \text{0.07}\\
    
    \addlinespace[2ex]   
    \bottomrule
\end{tabular*}

\end{threeparttable}
\vspace{-4mm}
\end{table}

\section{Conclusion}
We proposed the first network that simultaneously leverages both first and second order statistics for pooling in spatial and channel-wise attention mechanisms, resulting in a lightweight and fast model that restores high frequency image details. We compared our proposed model with various traditional and DL-based SR techniques on three CLE datasets in terms of image quality assessment metrics and inference time. Our RBAM network outperforms existing lightweight methods across different datasets, downsampling factors, and SR performance evaluation criteria. Experimental results also highlight the potential applicability of inexpensive software-based post-processing SR modules that improve degraded images in miniaturized CLE devices in real-time. \par


\noindent \textbf{Acknowledgments}. Thanks to the NVIDIA Corporation for the donation of Titan X GPUs used in this research and to the Collaborative Health Research Projects (CHRP) for funding.
\bibliographystyle{IEEE}
\bibliography{refs}

\clearpage
\end{document}